\documentclass[11pt]{article}
\usepackage{graphicx,url,amssymb,amsmath,rotating,color,units,wasysym,epsfig,multirow,epstopdf,stackengine}
\usepackage{comment}
\usepackage{cancel}
\usepackage{physics}
\usepackage{appendix}
\usepackage{mathtools}
\usepackage[colorlinks,urlcolor=blue,citecolor=blue,linkcolor=blue]{hyperref}
\usepackage{cleveref}
\usepackage[dvipsnames,svgnames]{xcolor}
\usepackage[normalem]{ulem}
\usepackage{tabularx}
\usepackage{enumitem}
\usepackage{tensor}
\usepackage{subcaption}
\usepackage{booktabs,makecell,float}
\usepackage{tikz}
\usetikzlibrary{arrows.meta, positioning}
\usepackage{pstricks}
\usepackage[utf8]{inputenc}
\usepackage{microtype}
\usepackage{soul}
\usepackage[commentmarkup=uwave,commandnameprefix=always]{changes}
\usepackage{tikz}
\usetikzlibrary{decorations.pathmorphing}
\tikzset{snake it/.style={decorate, decoration=snake}}
\def\ba{\begin{eqnarray}}
\def\ea{\end{eqnarray}}
\def\be{\begin{equation}}
\def\ee{\end{equation}}

\newcommand{\e}{{\rm e}}

\usepackage{xcolor}

\newcommand\bseq{\begin{subequations}}
\newcommand\eseq{\end{subequations}}

\newcommand{\bea}{\begin{eqnarray}}
\newcommand{\eea}{\end{eqnarray}}

\DeclareUnicodeCharacter{1F44D}{\thumbsup}

\definechangesauthor[name=SG, color=blue]{SG}
\definechangesauthor[name=LB, color=violet]{LB}
\definechangesauthor[name=LL, color=purple]{LL}

\usepackage{xcolor} % in the preamble

\usepackage[style=numeric-comp, backend=biber, sorting=none]{biblatex}
\numberwithin{equation}{section}  % Resets equation number at each section
\usepackage{float}
\begin{document}
\begin{titlepage}
\clearpage

\title{
{\bf A bulk model for dissipation and fixing EFT equations on its boundary}}
\author{A. Besharat$^{a,b}$\footnote{abeshara@ualberta.ca}, \,  L. Lehner$^{b}$\footnote{llehner@perimeterinstitute.ca},  \, J. Radkovski$^{b,c}$\footnote{jradkovski@perimeterinstitute.ca}~\\[2mm]
{\small\it $^a$ Department of Physics, University of Alberta, Edmonton, Alberta T6G 2J1, Canada,}\\  
{\small\it $^b$Perimeter Institute for Theoretical Physics, 31 Caroline St N,  Waterloo, ON N2L 2Y5, Canada} \\
{\small\it $^c$ Department of Physics and Astronomy, McMaster University,}\\
{\small \it 1280 Main St W, Hamilton, ON L8S 4M1, Canada}
}
\date{}

\maketitle

\begin{abstract}
Effective field theories (EFTs) with higher derivatives often yield equations of motion that define ill-posed problems. We present an action-based approach that supplements a recent well-posed formulation of Figueras et al. with selective dissipation: the EFT field is coupled to an auxiliary field living in one extra dimension, into which the energy of short-wavelength modes can dissipate. The construction is tested numerically on a scalar EFT with a known ultraviolet completion. We observe a reduction in spurious oscillations, and, for large couplings, in error. We also compare with the fixing-the-equations approach, and outline the extension to gravity.
\end{abstract}

\thispagestyle{empty}
\end{titlepage}

\newpage

\tableofcontents

\section{Introduction}
%################################################################
Effective field theory (EFT) is a powerful approach to study the low-frequency (macroscopic) dynamics of systems admitting separation of scales (see \cite{Weinberg1995, Burgess2020} for a textbook treatment). It is a particularly useful formalism when the complete microscopic description of the system is either unknown, or known, but too complicated. In the context of field theory, one typically uses EFT techniques to construct actions for the low-energy degrees of freedom order by order in the derivatives of fields. At each order, the particular choice of terms is dictated by the symmetries, analyticity and other assumed general properties of the system under study. The terms with fewest derivatives produce the leading contribution to the low-energy observables, whereas the higher-derivative terms provide corrections induced by the -- known, or unknown -- high-energy degrees of freedom.

The higher-derivative EFT corrections typically come suppressed by some high-energy scale and therefore, by assumption, are \textit{small}. In particular, in the context of gravitational physics, the leading term is typically given by the Einstein-Hilbert action for General Relativity (GR), and corrections are provided by higher-curvature terms, built out of the Riemann tensor, such as $R_{\mu \nu \sigma \rho} R^{\mu \nu \rho \sigma}$, $R_{\mu \nu \rho \sigma} R^{\mu \nu \lambda \xi} R^{\rho \sigma }_{\lambda \xi}$ and so on \cite{Donoghue1994, Donoghue2017}. With some ultraviolet (UV) completion of GR in mind, one typically expects corrections to be suppressed by the Planck mass, rendering them essentially unobservable. Importantly, this need not always be the case. One familiar example is Ho\v rava-Lifshitz Gravity \cite{Horava2008, Horava2009}. Here the suppression is provided by the much lower (but still very large) Lorentz-violating scale (see  the review \cite{Herrero-Valea2023})\footnote{Note that this theory is \textit{not}, strictly speaking, the UV completion of GR: Lorentz violation affects the leading-order dynamics.}. Generally, whatever the complete UV theory might be, as long as it predicts testable corrections, one can probe the induced \textit{classical} strong-gravity effects as deviations from predictions within General Relativity. Of particular relevance are gravitational waves which access the strong field/highly dynamical regime. This regime is potentially the one with the highest chance of inducing deviations (see e.g. \cite{LIGOScientific:2021sio,Krishnendu:2021fga}, and the recent GWTC-4.0 tests \cite{LIGOScientific:2026qni}), assuming, crucially, the scale of new physics is within reach of observations.
The prediction of signatures in specific theories, or frameworks to
capture deviations from General Relativity has been the focus
of significant attention in recent years (for some representative examples
see e.g.~\cite{Barausse:2012da,Bernard:2019fjb,Endlich2017,Loutrel:2022tbk,Corman:2022xqg,Cayuso:2023xbc,Schumacher:2023cxh,Julie:2024fwy,Bernard:2025dyh,Horowitz:2023xyl}).

Immediately, however, one encounters a formal problem. While corrections to GR are assumed to be small, they yield equations of motion that often spoil the hyperbolicity found in GR. This, in particular, can lead to instabilities and acausal behaviour (e.g.~\cite{Cayuso:2017iqc,Ripley:2019hxt,Bernard:2019fjb}). In a linear regime, a formal cut-off can be implemented to ensure effects beyond a certain scale are not at play, but this is more difficult in the non-linear regime. Such a regime typically requires simulations and these formal problems become pressing. The discretization of the equations introduces high-frequency numerical features lying outside the regime of validity of EFT. These features generically induce instabilities or, even worse, lead to the very demise of the initial value problem, e.g.~\cite{Ripley:2019hxt,Bernard:2019fjb,Franchini:2022ukz}.

\textit{Three} strategies have been proposed for dealing with these issues. The first, most straightforward, approach is to work order by order in derivatives solving the equations of motion iteratively. This procedure yields spurious behavior of solutions due to secular effects \cite{Allwright:2018rut,Cayuso:2020lca,Corman:2024cdr,Heller:2025dxh}. While these might not be significant in scattering processes, they can easily spoil predictions when long interactions occur. An example is
the gravitational inspiral and coalescence in a binary problem.  The second approach is to introduce some \textit{new} fields that (partially) bring back heavier modes which had been integrated out, but in a way that they ``dissipate away'' the energy contained in, or cascading toward, the UV beyond the EFT regime. In the context of hydrodynamics, a classical example of this approach is the M\"uller--Israel--Stewart (MIS) theory \cite{Muller:1967zza, Israel:1976tn, Israel:1976efz, Israel:1979wp, Hiscock:1983zz, Baier:2007ix, Denicol:2012cn}\footnote{Originally formulated at the level of the constitutive relations, it has recently been recast into the Schwinger-Keldysh EFT with the addition of noise \cite{Jain2023}.}.  In gravitational physics the analogue is the \textit{fixing-equations} approach of \cite{Cayuso:2017iqc}. This approach is implemented at the level of the equations of motion, and the new fields break the covariance of the theory once dissipation is introduced.
Finally, a third approach is to conveniently redefine the \textit{existing} fields so that the equations of motion in terms of the redefined fields have a particular character that enables one to establish well-posedness at the analytical level. 
In the context of dissipative relativistic hydrodynamics, such a method -- BDNK hydrodynamics -- was implemented at the level of the constitutive relations  \cite{Bemfica2017, Kovtun2019, Bemfica2019, Hoult2020, Bemfica2020}. Its relative in gravity \cite{Figueras:2024bba, Figueras:2025gal} does indeed address the most delicate pathologies: while the resulting equations of motion are only weakly hyperbolic, they have a structure that ensures well-posedness can be achieved. The lack of strong hyperbolicity of the system, though, comes at a cost at the practical level. The rigorous applied mathematics theory that guarantees a consistent and stable
implementation preserving key properties of the continuum problem does not apply~\cite{KreissLorenz1989,gustafsson2013time}, and the consistent treatment of boundaries, especially in general constrained systems, is not fully understood. Thus, one launches the numerical enterprise from less firm ground.

One appealing feature of the method \cite{Figueras:2024bba} is its implementation at the level of the action. Although the action is practically indispensable for computing \textit{quantum} observables, in \textit{classical physics} one in principle only needs the equations of motion. Still, it serves as an interface with quantum physics and as a ``repository'' for the symmetries of the problem. At a practical level, an action also makes perturbative approaches systematic and efficient \cite{Goldberger2006eft, Porto2016eft}. In addition to that, having an action enables symplectic/variational integrators in numerical applications~\cite{MarsdenWest2001, Stern_2015} (see \cite{Galley:2012hx,Tsang2015} for the approach to solving non-conservative mechanical dynamics using an action of the Schwinger-Keldysh type). Importantly, action encodes field-redefinition freedom, which is precisely what was exploited by \cite{Figueras:2024bba} to modify the theory in a consistent way so as to control its behavior in the UV and ensure well-posedness.

Building on \cite{Figueras:2024bba}, we bring to the level of the action the ideas of the fixing method \cite{Cayuso:2017iqc}. We represent the EFT fields as boundary values of fields defined in one extra dimension \cite{Besharat2023}. The full boundary-plus-bulk system is conservative and, with bulk operators chosen appropriately, allows one to consistently describe the dissipative dynamics on the boundary. The general formalism allows one to model the ``fixing'' dissipation that carries away the energy from the high-frequency modes into the extra dimension in the form of a flux. At the same time, the framework is flexible enough to describe real, physical dissipation and -- when put on the Schwinger-Keldysh contour -- the corresponding noise. Our goal in this work is to demonstrate how in practice this setup can bring more control over the resulting dynamics outside the EFT regime. We construct an appropriate bulk action, integrate out the bulk degrees of freedom analytically and then solve the problem for the original fields. We focus on the simple case of a scalar field theory and illustrate the corresponding gravity construction. A thorough analysis for gravity, as well as the implementation through variational integrators, is left for future work. To summarize, we enhance the approach presented in \cite{Figueras:2024bba} by including an optional dissipative mechanism, and, at the same time, we adapt the ideas introduced in \cite{Cayuso:2017iqc} to implement the mechanism through a more flexible, action-based methodology.

The structure of this work is as follows. We present the method in Sec. \ref{sec:fixing_action} through the example of an EFT for a self-interacting scalar field in $3+1$ dimensions. Following \cite{Figueras:2024bba} we use the field-redefinition freedom and consider the scalar field as living on the boundary of a space extended by one auxiliary dimension. We then write a theory in the boundary-plus-bulk introducing an auxiliary field, playing the role of the dissipative degrees of freedom in the MIS-like approaches. We show that with an appropriate boundary condition, this action reproduces the original classical dynamics for the scalar augmented with the damping terms. Finally, we discuss the generalization to gravity. In Sec. \ref{sec:numerics} we illustrate the resulting behavior, compare with results of the UV-complete theory (see also \cite{Allwright:2018rut,Figueras:2025gal}) and show that the damping terms give further control over the solution in the UV \cite{Cayuso:2017iqc,Allwright:2018rut}. We conclude in Sec. \ref{sec:conclusion} with a discussion of the results and the outlook. In Appendix \ref{appendix} we present an in-in formulation that encodes the noise.
 
\section{Fixing the action}
\label{sec:fixing_action}
\subsection{Bulk model}
Our ultimate goal is to turn the ill-posed conservative dynamics with higher-derivative terms into well-posed dynamics with selective dissipation combining the logic of~\cite{Figueras:2024bba} and \cite{Cayuso:2017iqc}. We start by introducing the dissipation at the level of the action for a theory that gives an equation of motion naturally leading to a well-posed problem (containing no higher-derivative terms). The necessary modification for the higher-derivative case is straightforward. Consider for concreteness the action for a free massless scalar in $3+1$ dimensions. The construction will not depend on the dimension of the problem, and in Sec.~\ref{sec:numerics} we in fact consider the simpler $1+1$ case for the numerical implementation. We will display elementary derivations to set up the notation. Assuming the signature $\eta_{\mu \nu} = \text{diag}(-1,1,1,1)$ and coordinates $x^{\mu} = (t, \mathbf{x})$, we write
\begin{eqnarray}
\label{eq:actionphitwo}
    S_0 = -\int d^4x \, \frac{1}{2} (\partial_{\mu} \phi)^2 \, .
\end{eqnarray}
Trivially we can find the equations of motion by varying the action,
\begin{eqnarray}
\label{eq:variation_S0}
    \delta S_0 = \int d^4x \, \delta \phi \, \Box \phi \, , \quad \Box \equiv \partial_{\mu}\partial^{\mu} = - \partial_t^2 + \partial_{\mathbf{x}}^2 \, ,
\end{eqnarray}
and using the variational principle, from which it follows,
\begin{eqnarray}
\label{eq:conservative_dynamics}
    \Box \phi = 0 \, .
\end{eqnarray}
We now show that we can promote the classical conservative dynamics to the dissipative one,
\begin{equation}
    \Box \phi =0 \quad \rightarrow \quad  \Box \phi - \tau \partial_t \phi = 0 \, ,
\end{equation}
with $\tau>0$ to ensure that the field \textit{loses} energy by coupling it to a reservoir (see, e.g.~\cite{1986GReGr..18..913K}). Namely, we attach to the field $\phi$ a ``string'' at every spatial point $\mathbf{x}$. More specifically, the reservoir can be understood as a bath of non-interacting oscillators parametrized by the coordinate $z$ \cite{Caldeira1982} and coupled to the field $\phi(t,\mathbf{x})$ at each point $\mathbf{x}$. The basic idea is in fact \textit{very} old and first appeared in \cite{Lamb1900} (in the context of dissipative mechanics).

Explicitly, we add into the theory a free dynamical field $\chi$ that depends on an auxiliary dimension $z>0$ as well as on $x^{\mu}$,
\begin{eqnarray}
\label{eq:action_diss}
    S_0 \rightarrow S = S_0 +  \int_{z>0} d^4x \, dz \, \frac{\tau}{2}\Big( (\partial_{t} \chi)^2 -  (\partial_{z} \chi)^2 \Big) \, ,
\end{eqnarray}
with the two fields $\phi$ and $\chi$ related by the boundary condition
\begin{eqnarray}
\label{eq:boundary_condition}
    \chi(z=0,t,\mathbf{x}) = \phi(t,\mathbf{x}) \, .
\end{eqnarray}
Importantly, in \eqref{eq:action_diss} we do not add a mass term for the field $\chi$.
Neither do we add any derivatives with respect to the spatial coordinates $\mathbf{x}$: from the perspective of the dynamics in the $(t,z)$-plane these also appear as the mass terms.\footnote{That is, after the Fourier transformation in $\mathbf{x}$, they would look like a mass term with mass proportional to the Fourier momentum.} A non-zero ``$(t,z)$-mass'' means reservoir modes are gapped, and correspondingly the boundary modes with frequency below the gap would be totally reflected instead of absorbed. Given that, this soon-to-be dissipative theory is \textit{not} effective, as we do not write all the possible terms.

Varying the action $S$, we have
\begin{align}
\label{eq:variation_S}
    \delta S &= \delta S_0 + \tau \int_{z>0} d^4x \, dz \, \delta \chi \Big( \partial_z^2 - \partial_t^2\Big) \chi \nonumber \\
    &+ \tau  \int_{z>0} d^4x \, dz \, \partial_{t} (\delta \chi \partial_t \chi) - \tau  \int_{z>0} d^4x \, dz \, \partial_{z} (\delta \chi \partial_z \chi) \, .
\end{align}
Here the first term $\delta S_0$ is given by \eqref{eq:variation_S0}. The second term is obtained by varying the action for $\chi$ \textit{and} integrating by parts. It gives the (wave) equation of motion for $\chi$. Importantly, we keep the boundary terms in the second line.

With no mass terms or gradient terms in $\mathbf{x}$, the equation of motion for $\chi$ admits wave solutions $\chi(z,t,\mathbf{x}) = \chi(t\pm z, \mathbf{x})$. To obtain the dissipative dynamics we take the field $\chi$ to be purely \textit{outgoing}
\begin{eqnarray}
\label{eq:outgoing}
    \chi(z,t,\mathbf{x}) \equiv \bar{\chi}(t-z, \mathbf{x}) \, , \quad \bar{\chi}(t,\mathbf{x}) = \phi(t,\mathbf{x}) \, ,
\end{eqnarray}
as well as vanishing at infinity,
\begin{eqnarray}
\label{eq:vanishing_at_inf}
    \bar{\chi}(\pm \infty, \mathbf{x}) = 0 \, .
\end{eqnarray}
This agrees with the physical intuition that the field $\chi$ takes the energy away from the system described by $\phi$.

Having imposed the conditions above, we are left only with the first and the last terms in $\delta S$. Using, in order, the conditions \eqref{eq:vanishing_at_inf}, \eqref{eq:outgoing} and the boundary condition \eqref{eq:boundary_condition}, we write for the last term in \eqref{eq:variation_S}:
\begin{align}
     -\tau  \int_{z>0} d^4x \, dz \, \partial_{z} (\delta \chi \partial_z \chi) &= + \tau  \int d^4x  \, \delta \chi \partial_z \bar{\chi} \biggr|_{z=0} \nonumber \\
    &= - \tau  \int d^4x  \, \delta \chi \partial_t \bar{\chi} \biggr|_{z=0} = -\tau  \int d^4x  \, \delta \phi \partial_t \phi \, .
\end{align}
Assembling everything together we arrive at the sought-after equation of motion for $\phi$ with dissipation
\begin{eqnarray}
\label{eq: dissequation}
    \Box \phi - \tau \partial_t \phi = 0 \, .
\end{eqnarray}
The resulting equation of motion now describes a field that is dissipated on a timescale given by $\tau^{-1}$.

\subsection{The scalar EFT and selective dissipation}
\paragraph{The UV model.}
The dissipative mechanism in \eqref{eq:action_diss}, however, does not target the short-wavelength modes. We next make the dissipation selective, using, as a setting, a theory whose equation of motion does not define a well-posed problem: the model of~\cite{Burgess:2014lwa} (see also~\cite{Allwright:2018rut}), a complex scalar field with a Mexican-hat potential,
\begin{eqnarray}
    S_{\rm UV} = - \int d^4 x \left( \partial_{\mu} \varphi^* \partial^{\mu} \varphi + \frac{M^2}{2 v^2} \left(\varphi^* \varphi - \frac{v^2}{2} \right)^2  \right) \, .
\end{eqnarray}
If one is interested in the dynamics at long time scales, a simpler, effective, theory can be derived by integrating out the massive fluctuations around the vacuum. After expanding around the symmetry-breaking vacuum via
\begin{equation}
    \varphi = \frac{v}{\sqrt{2}} \bigl(1+\rho\bigr)e^{i\phi} \, ,
\end{equation}
the action for the radial mode $\rho$ and the Goldstone mode $\phi$ becomes \footnote{After rescaling with $v$.}
\begin{equation}\label{eq:Suv}
S_{\rm UV} \rightarrow -\int d^4 x \, \left( \frac{1}{2}(\partial_{\mu} \rho)^2
+ \frac{1}{2}(1+\rho)^2 (\partial_{\mu} \phi)^2
+ V(\rho) \right)\, ,
\end{equation}
with
\begin{equation}
V(\rho) = \frac{M^2}{2}\left(\rho^2 + \rho^3 + \frac{\rho^4}{4}\right) \, .
\end{equation}
The field $\rho$ is a massive field of mass $M$, while $\phi$ is a massless Goldstone boson.

In the low-energy regime $E \ll M$, the heavy field $\rho$ can be integrated out, yielding an effective field theory for $\phi$. To leading nontrivial order, the effective Lagrangian is
\begin{equation}
\label{eq:Seffmain}
S_{\text{eff}}
= -\int d^4x \left( \frac{1}{2}(\partial_{\mu} \phi)^2
- \frac{\gamma}{4} (\partial_{\mu} \phi)^4
+ \mathcal{O}(M^{-4}) \right) \, ,
\end{equation}
with $\gamma \equiv 2/M^2$. This higher-derivative interaction encodes the effects of the heavy mode and introduces nonlinear derivative self-interactions in the effective dynamics.

This model has been the subject of several works discussing the dangers (and ways to avoid them) of employing EFTs beyond their regime of validity, see e.g.~\cite{Burgess:2014lwa,Allwright:2018rut,Figueras:2025gal}. Particularly appealing is the fact that the complete model is well-behaved in the UV, and its solutions can be used to assess the quality of the EFT-derived solutions.

Since at leading order the effective action \eqref{eq:Seffmain} yields equation of motion $\Box \phi = 0$, we are at liberty to consider a field redefinition, following \cite{Figueras:2024bba},
\begin{eqnarray}
\label{eq:field_redefinition}
    \phi \rightarrow \phi - \frac{\alpha}{2} \Box \phi \, .
\end{eqnarray}
Parameter $\alpha$ can be chosen at will, and in particular it can always be taken positive. After dropping the higher order terms we get a modification of the action,
\begin{align}
\label{eq:action}
S_0 = -\int d^4x \left( \frac{1}{2}(\partial_{\mu} \phi)^2 - \frac{\gamma}{4}(\partial_{\mu} \phi)^4 - \frac{\alpha}{2} \partial_{\mu}\phi \partial^\mu \Box\phi \right) \, ,
\end{align}
and the corresponding equation of motion,
\begin{align}
\label{eq:eq_of_motion}
\Box \phi = \gamma \Box \phi (\partial_{\mu} \phi)^2 + 2\gamma \partial_{\mu}\partial_{\nu}\phi\, \partial^{\mu}\phi\, \partial^{\nu} \phi + \alpha\, \Box \Box \phi \, .
\end{align}
In the spirit of the EFT approach, the last term in \eqref{eq:action} -- and hence in \eqref{eq:eq_of_motion} -- is redundant: we put it in by hand using a field redefinition \eqref{eq:field_redefinition}. It only alters the dynamics precisely where the EFT no longer applies, yet it is kept by \cite{Figueras:2024bba} for a reason. Without it, solutions can lose hyperbolicity at high frequencies, rendering the initial-value problem ill-posed, but generic numerical implementations would feed modes with such frequencies, regardless of the physical situation in mind. With it, the principal part of the equation is the fixed operator $\alpha\Box^2$. The equation is then only weakly hyperbolic ($\Box^2$ has degenerate eigenvectors), but the lower-order structure -- the $\Box$ operator in \eqref{eq:eq_of_motion} -- still allows (local) well-posedness to be established. The modified system can thus be regarded as an ``effective UV completion'' that reproduces the original problem within the EFT regime. Of course, the question remains as to the consequences of not working with a strongly hyperbolic system and not playing ``by the rules of EFT'' by bringing in the redundant operator (and, related, whether this addition can affect the physics one is interested to study). Importantly, from  a physics standpoint, to examine the behavior of the system, the transformation inverse to that of \eqref{eq:field_redefinition} needs to be performed to recover the behavior of the original theory (corresponding to $\alpha=0$). For weak couplings, the effect of this redefinition can be negligible, becoming relevant outside the EFT regime to control short-wavelength modes. In practice, for general problems, such inversion need not be straightforward or efficient to perform. In that view, the dissipative term introduced next provides a ``control knob'' with which potential negative effects can be removed or at least ameliorated.
%%%%%%%%%%%%%%%%%%%%%%%%%%%%%%%%%%%%%
\paragraph{Selective dissipation.}
As emphasized, the dissipative term of the form \eqref{eq: dissequation} is not suitable here: it damps all modes equally, including the long-wavelength ones that should follow the leading-order dynamics $\Box\phi=0$. To ``fix'' the high-frequency modes only, we extend the conservative action $S_0$, choosing $\tau>0$, by the following dissipative term: \footnote{The parameter $\tau$ in \eqref{action+} of course carries different dimensions than the one in the model \eqref{eq:action_diss}, but we choose to keep the same notation.}
\begin{eqnarray}
\label{action+}
S_0 \, \rightarrow \, S  &=& -\int d^4x \left( \frac{1}{2}(\partial_{\mu} \phi)^2 - \frac{\gamma}{4}(\partial_{\mu} \phi)^4 - \frac{\alpha}{2} \partial_{\mu}\phi \partial^\mu \Box\phi \right) \nonumber \\
&& + \int_{z>0} d^4x \, dz \,\frac{\tau}{2}  \,\big((\partial_{t} \partial_{\mu} \chi )(\partial_{t} \partial^{\mu} \chi)-\partial_{z} \partial_{\mu} \chi \partial_{z} \partial^{\mu} \chi \big) \, ,
\end{eqnarray}
We impose the boundary condition $\chi|_{z=0}=\phi$ as before and keep the purely outgoing solution, $\chi(z,t,\mathbf{x})=\phi(t-z,\mathbf{x})$.\footnote{The equation in the bulk is $\Box(\partial_t^2-\partial_z^2)\chi=0$. We take $\chi=\phi(t-z,\mathbf{x})$, which is a solution with no extra $\Box$-homogeneous piece.} Going through the same boundary variation as for \eqref{eq:action_diss} then gives
\begin{align}
\label{eq of mo2}
\Box \phi = \gamma \Box \phi (\partial_\mu \phi)^2 + 2\gamma (\partial_{\mu} \partial_{\nu} \phi)\, \partial^{\mu}\phi\, \partial^\nu \phi + \alpha\, \Box \Box \phi - \tau\, \partial_t \Box \phi \;.
\end{align}
The choice \eqref{action+} is not unique and should be determined on a case-by-case basis from the requirement of yielding correct dissipation. We do not attempt to consider the most general case here.\footnote{A particular alternative choice is to work with four ``strings'', i.e. fields $\chi_{\mu}$, with dissipative action,
\begin{eqnarray}
    \int d^4x \, dz \,\frac{\tau}{2}  \big(\partial_{t}\chi_{\mu} \partial_{t}\chi^{\mu}-\partial_{z}\chi_{\mu} \partial_{z}\chi^{\mu} \big) \, ,
\end{eqnarray}
and boundary condition,
\begin{equation}
    \chi_{\mu}(z=0, t, \mathbf{x}) =  \partial_{\mu} \phi(t, \mathbf{x}) \, .
\end{equation}
The resulting dynamics after ``integrating out'' fields $\chi_{\mu}$ is given by the same equation \eqref{eq of mo2}. For other theories a different choice could be more appropriate.} 

In Appendix \ref{appendix} we examine the question of whether one can arrive at this dynamics from an interaction between the system described by $\phi$ and some bath. We take the case of a weakly coupled hot thermal bath and show that one can indeed obtain the equation \eqref{eq of mo2} by placing the theory on a Schwinger-Keldysh time contour \cite{Bakshi19621, Mahanthappa19622, Bakshi1963, Keldysh1964} and integrating out the bath in the path integral formalism. The resulting action with doubled fields $\phi_{\pm}$, corresponding to two branches of the contour, has the dissipative equation \eqref{eq of mo2} for the ``average'' field $\frac{\phi_{+} + \phi_{-}}{2}$ as its saddle point. A seemingly unavoidable issue with this procedure is the need for a bath with a spectral density that is not positive definite for all the bath modes. This is expected, since the ``oscillators'' $\chi$ have a fourth-order action in \eqref{action+}.

The impact of the term proportional to $\alpha$ and the dissipative term in \eqref{eq of mo2} can be analyzed by considering (for simplicity) the perturbations of the constant solution,
\begin{equation}
    \phi = \mbox{const} + \delta \phi \, ,
\end{equation}
and using Fourier analysis to determine the dispersion relations of the perturbation modes. At linear order in $\delta \phi$ we have
\begin{equation}
\label{eq: linear}
 (1-\alpha \Box + \tau \partial_t)\Box \delta \phi  = 0 \, .
\end{equation}
It is straightforward to see that this equation propagates plane waves corresponding to the leading order equation and the modes decaying in time, ensuring that the long-wavelength solutions are those determined by the leading-order dynamics, consistently with the EFT assumptions. The local well-posedness argument of~\cite{Figueras:2024bba} applies directly for $\tau>0$.

Importantly, the dissipative action \eqref{action+} is not unique. For solutions that remain within the EFT regime the particular choice should not matter (see the discussion in e.g.~\cite{10.1063/1.530958}); if it did, that would mean data chosen within the EFT regime can evolve out of it, cascading toward the UV, and describing such dynamics would require the UV-complete theory or an extension of the EFT by resummation or related methods, e.g.~\cite{Heller:2025dxh}. We chose the form \eqref{action+} because it attaches the damping to the propagating modes; a sensitive dependence on this choice would therefore signal the breakdown of the EFT-derived description. 
\subsection{Extension to gravity}
Here we provide a formal illustration of how one can introduce such a dissipative mechanism in gravity. We do not analyze the gauge fixing, constraint propagation, or nonlinear well-posedness. As an example, take a first putative higher-derivative correction to GR (use the Gauss-Bonnet theorem plus field redefinitions to put it in this form),
\begin{equation}
S_0 = S_m + \frac{1}{2\kappa}\int\mathrm{d}^4x\,\sqrt{-g}\left[R-\alpha \,R_{\mu\nu}G^{\mu\nu}\right] \, , \quad  G_{\mu \nu} = R_{\mu \nu} - \frac{1}{2} g_{\mu \nu} R \, ,\label{beyondGR}
\end{equation}
where $S_m$ is some matter action that we leave unspecified. Matter is needed here: all correction and damping terms below are built from the Ricci tensor, so for $T_{\mu\nu}=0$ every vacuum solution of GR solves the modified equations exactly.
The idea is to simply follow the logic of the scalar construction above and extend the metric
into the bulk, defining $\gamma_{\mu\nu}(t,z,x)$ such that 
\begin{eqnarray}
    \gamma_{\mu \nu} \Bigr|_{z=0} = g_{\mu \nu} \, .
\end{eqnarray}
We can then mimic the action \eqref{action+} by using derivatives of the metric, i.e. Christoffel symbols, instead of $\partial_{\mu} \chi$,
\begin{equation}
    \Gamma^{\mu}_{\nu \sigma} = \frac{1}{2} \gamma^{\mu \rho} \left(\gamma_{\nu\rho,\sigma}
          +\gamma_{\sigma\rho,\nu} 
          -\gamma_{\nu\sigma,\rho}\right) \, .
\end{equation}
A particular combination that yields correct dissipative dynamics is achieved by the action 
\begin{align}
S = S_0  -\frac{\tau}{2\kappa}\int_{z>0}d^4x \, dz\,\sqrt{-\gamma}\,
\Big[\langle \partial_t \Gamma,\partial_t \Gamma \rangle - \langle \partial_z \Gamma,\partial_z \Gamma \rangle \Big]
\, ,\label{gravdiss}
\end{align}
where to avoid clutter we define a shorthand ``scalar product'' on objects with index structure of Christoffel symbols, 
\begin{equation}
\langle V,V \rangle = V^{\mu}_{\nu\rho}\,\gamma^{\sigma\nu} \, V^{\rho}_{\sigma\mu}
-V^{\mu}_{\nu\rho}\gamma^{\nu\rho} \, V^{\sigma}_{\mu\sigma} \, .
\end{equation}
Up to a total derivative, this combination, when evaluated on Christoffel symbols, is simply the Ricci scalar,
\begin{equation}
    \sqrt{- \gamma }\, \langle\Gamma, \Gamma \rangle = \sqrt{-\gamma} \, R + \text{total derivative}\, .
\end{equation}
On the outgoing solution $\gamma_{\mu\nu}(t,z,x)=g_{\mu\nu}(t-z,x)$, variation of $g$ yields
\begin{equation}
G_{\mu\nu}-\alpha \,\mathcal H_{\mu\nu}
+\tau\left(\partial_t G_{\mu\nu}
+\tfrac12 R\,\partial_t g_{\mu\nu}
-\tfrac12 g_{\mu\nu}\,R^{\rho\sigma}\partial_t g_{\rho\sigma}\right)=\kappa T_{\mu \nu} \, , \label{extraterm}
\end{equation}
with
\begin{equation}
\mathcal H_{\mu\nu}=\Box R_{\mu\nu}-\tfrac12 g_{\mu\nu}\Box R
+2R_{\mu\rho\nu\sigma}R^{\rho\sigma}-R\,R_{\mu\nu}
-\tfrac12 g_{\mu\nu}R_{\rho\sigma}R^{\rho\sigma}+\tfrac14 g_{\mu\nu}R^2 \, ,
\end{equation}
and $T_{\mu\nu}$ the energy-momentum tensor corresponding to the matter action $S_m$. Linearizing around $g_{\mu\nu}=\eta_{\mu\nu}+h_{\mu\nu}$, with transverse-traceless $h_{\mu\nu} \equiv h^{\mathrm{TT}}_{\mu \nu}$, we get
\begin{equation}
\left(1 - \alpha \Box +\tau\,\partial_t\right) \Box h^{\mathrm{TT}}_{\mu\nu}=  -2\kappa T^{\text{TT}}_{\mu \nu} \, ,
\end{equation}
which is of the same form as \eqref{eq: linear} and therefore propagates only plane waves and the decaying modes. The action \eqref{beyondGR} is only an illustrative example. For any beyond-GR theory with higher-curvature corrections, adding \eqref{gravdiss} yields equations of motion with the same damping terms as in \eqref{extraterm} (only the expression for ${\cal H}_{\mu\nu}$ changes).
 
\section{Numerical Implementation}
\label{sec:numerics}
We now implement three different problems and compare their solutions.
The first one is the original problem in terms of $\{\phi,\rho\}$ of the UV-complete model in \eqref{eq:Suv} (see also \cite{Burgess:2014lwa}).
The second ($C_1$) is the EFT-derived one~\eqref{eq of mo2}, which includes both the dissipative contribution and the operator that could be field-redefined away. Finally, the third ($C_2$) corresponds to the same model~\eqref{eq of mo2}, but without the operator that can be redefined away, \textit{and} implemented following the fixing approach, as described in~\cite{Allwright:2018rut}. Briefly, in this approach \cite{Cayuso:2017iqc} the higher-derivative source is evolved indirectly: one solves $\Box\phi=\Pi$ with $\Pi$ an independent field obeying the equation (Eqs.~(27)--(28) of \cite{Allwright:2018rut}),
\begin{equation}
\label{eq:C2eq}
    \Pi = \gamma \, (\partial_\mu\phi)^2\,\Pi + 2 \gamma \,\partial_{\mu}\partial_{\nu}\phi\,\partial^{\mu}\phi\,\partial^{\nu}\phi - \tau \,\partial_t\Pi \, .
\end{equation}
Here $\tau>0$, and on this timescale the equation relaxes $\Pi$ to the value reproducing the EFT dynamics (and damps frequencies above $\tau^{-1}$).\footnote{Note that this parameter $\tau$ is a different object from the reservoir coupling $\tau$ of $C_1$.}

For simplicity we consider the 1+1 setting with periodic boundary conditions in the computational domain $x\in [0,10]$. We reduce the equations to first order in both time and space. We adopt
finite difference approximations with a uniform spatial grid employing a fourth-order approximation for spatial derivatives and a third-order Runge-Kutta scheme in time for the time evolution. We use
the method of lines with $\Delta t = 1/4 \,\Delta x$ and include the Kreiss-Oliger dissipation of order 6, e.g.~\cite{Calabrese_2003,Calabrese_2004}. Production runs have $\Delta x=L/(N-1) = 10/800$, though convergence tests and the studies of data with noise (as described below) are run at up to $10\times$ higher resolution, verifying that the expected third-order convergence is achieved. 
%%%%%%%%%%%%%%%%%%%%%%%%
\subsection{Initial data}
We define initial data for the field $\phi$ with compact support as
\begin{equation}
\label{eq:initdata}
\phi = A (x-x_L)^4 (x-x_R)^4/3.5^8 \quad {\rm if } \, x\in[x_L=2,x_R=5] \,  ,
\end{equation}
and $\phi =0$ otherwise, with amplitude $A$ fixed to $A=10$ throughout. The initial value for $\partial_x \phi$ is defined in accordance with \eqref{eq:initdata} and we set $\partial_t \phi = 0$.
When solving the full UV-complete problem, we follow the discussion provided in~\cite{Figueras:2025gal} to ensure that the initial data corresponds to the EFT regime. We also use this information to adapt the initial data for the fixing approach. Furthermore, to explore data with both noise \textit{and} ``constraint'' violations, we add some amount of low-/high-frequency noise by introducing 
\begin{equation}
\phi(t=0,x) = \phi(t=0,x) + A^2 \left( \sin(100 \pi x/L) + \cos(2 \pi x/L)  \right)/3.5^8 \, ,
\end{equation}
but leaving $\partial_x \phi$ intact. We confirm that in both cases $C_1$ and $C_2$, a stable evolution is achieved if one chooses values consistent with the EFT assumptions. Both approaches reliably
control the high frequencies and, in particular, $C_1$ does so even with $\tau=0$ (i.e. no dissipation added), as expected from a well-posed problem (see~\cite{Figueras:2024bba}). Thus, both approaches control the potential UV issues. In the next section, we concern ourselves with contrasting the obtained solutions with those provided by the UV-complete theory. Note that the comparison is direct, without performing the field-redefinition inversion.

\begin{figure}[t]
\centering
\includegraphics[width=9.2cm]{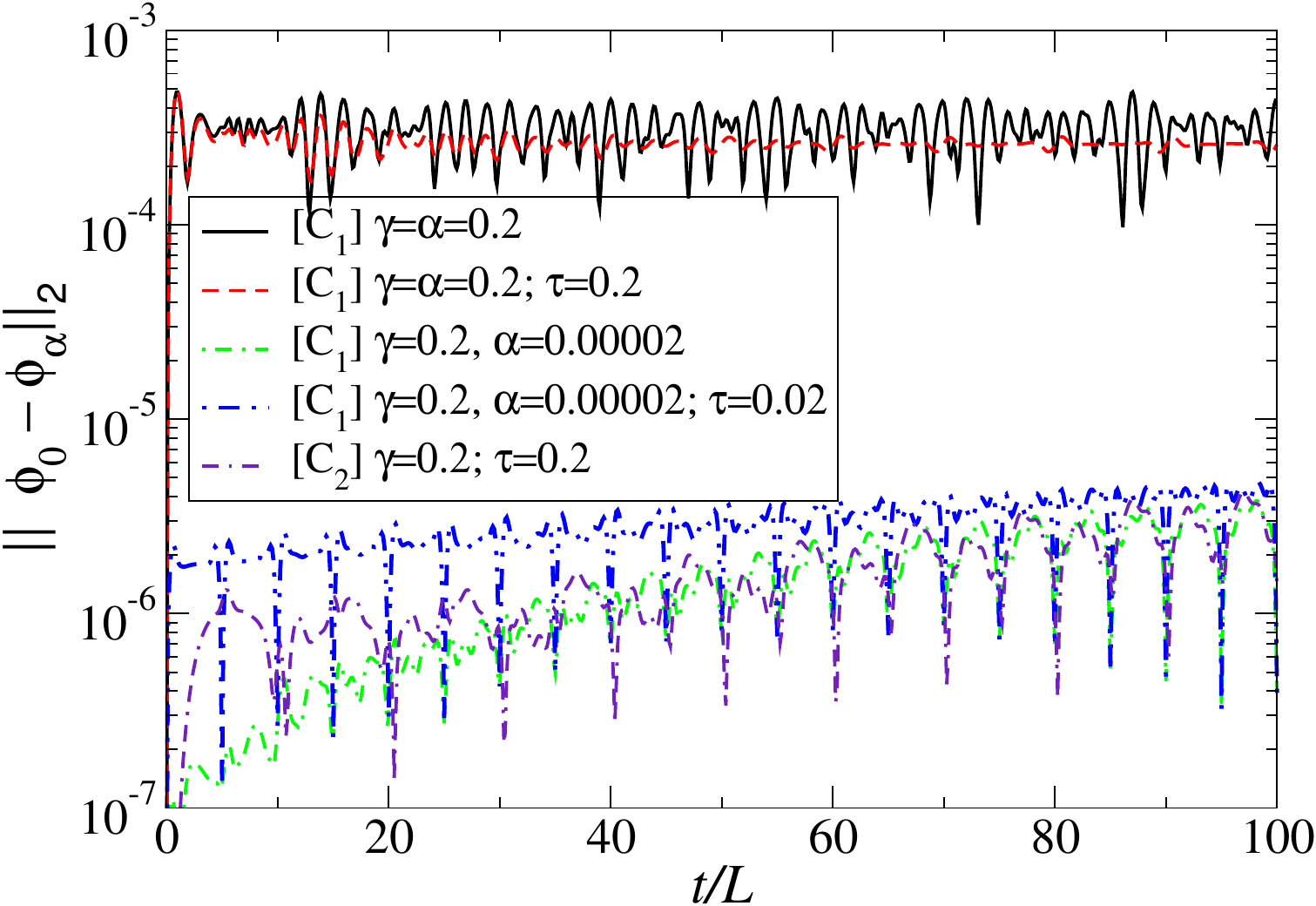}
\caption{ The $L_2$ norm of the differences between the $C_{1,2}$ solutions and the original problem at $\gamma=0.2$: $\phi_0$ for the UV-complete problem, $\phi_{\alpha}$ for the EFT. Parameter $\tau$ for $C_2$ is the one from \eqref{eq:C2eq}. A stable evolution is found for all methods. The smallest errors are obtained in the fixed method $C_2$ and with $\alpha \ll \gamma$ for $C_1$. The larger initial errors relative to the case $\tau=0$ for $\gamma=0.2$ are sourced by the contribution of the dissipative term.}
\label{smallamplitude}
\end{figure}
\subsection{Results}
We begin by considering a small value of the coupling $\gamma=0.2$. We then solve for values $\alpha=\{0.2 \, , 0.00002\}$  together with both zero and non-zero values of $\tau$ ($\tau=0.2$ for $\alpha=\gamma$ and $\tau=0.02$ for $\alpha\ll\gamma$) (Fig.~\ref{smallamplitude}). For this value of~$\gamma$ all cases can be run for the time specified (100 crossing times, to illustrate the expected long-time controlled behavior). A comparison with the solution of the UV-complete problem reveals that while all options give rise to stable behavior, the  discrepancies with the true solution are about two orders of magnitude larger when $\alpha=\gamma$ than for $\alpha \ll \gamma$. That is, when the higher order operator is suppressed by a heavier scale (smaller $\alpha$), its influence on the observed dynamics is less severe, as the further modes introduced by its inclusion stay essentially clamped. Introducing a damping mechanism helps to reduce spurious oscillations but overall the magnitude of errors is similar. The fixing approach provides an error similar to the case with the higher order operator suppressed by the heavier scale ($\alpha \ll \gamma$). This behavior is not surprising, as
the field introduced by the fixing approach is forced to stay faithful to the original EFT dynamics within some reasonably short timescale.

\begin{figure}[t]
\centering
\includegraphics[width=9.2cm]{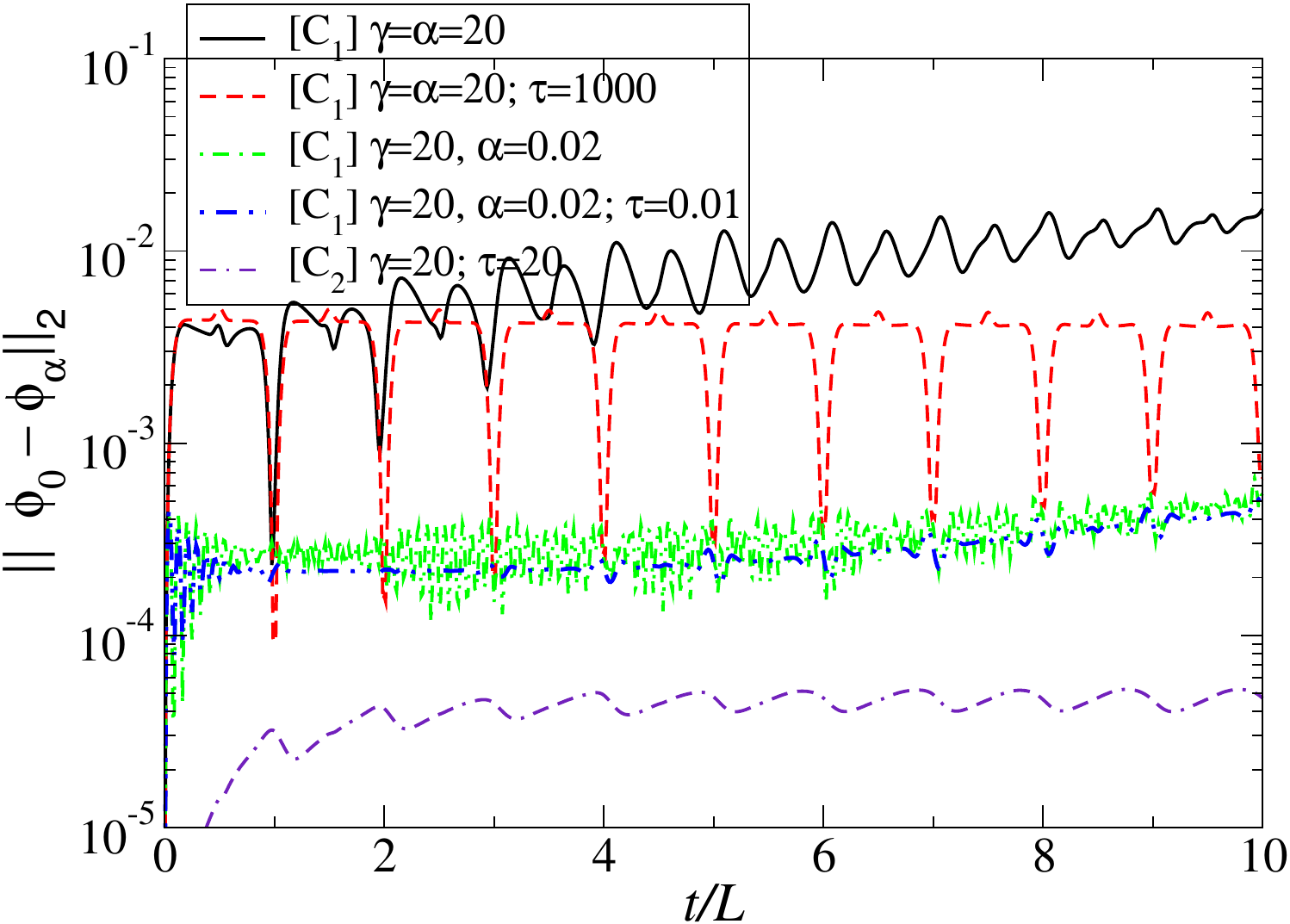}
\caption{The $L_2$ norm of the differences between the $C_{1,2}$ solutions and the original problem at $\gamma=20$.  The smallest errors are obtained in the fixed method $C_2$ and with $\alpha \ll \gamma$ for $C_1$. The addition of the damping reduces the differences with respect to the full solution, most notably for values of $\alpha\approx \gamma$.}
\label{largeamplitude}
\end{figure}
Next we choose $\gamma=20$ and evolve for 10 crossing times, with $\alpha=\{20,\,0.02\}$ and again both zero and non-zero values of $\tau$ ($\tau=1000$ for $\alpha=\gamma$, $\tau=0.01$ for $\alpha\ll\gamma$), see Fig.~\ref{largeamplitude}. This coupling brings the corrections perilously close to the boundary of the EFT regime.\footnote{Indeed, even a slightly larger value of $\gamma$ breaks hyperbolicity in a short time in the original problem with $\alpha=\tau=0$.} It is also the case of practical interest: observable differences from the leading-order theory require large couplings -- in the beyond-GR context, a new-physics length scale of order $\approx \text{km}$, within reach of gravitational-wave observations, rather than the Planck length. As before, the discrepancies with the full solution are largest for $\alpha=\gamma$; the damping now reduces both the spurious oscillations and the error (more so for $\alpha \sim \gamma$). For $\alpha\ll\gamma$ the errors decrease by one to two orders of magnitude (from $\sim10^{-2}$ to a few~$\times10^{-4}$ in Fig.~\ref{largeamplitude}), and further with the fixing approach. Importantly, the practical limitations make it costly to reduce $\alpha$ much further, as the equations become stiff. This, in turn, taxes the numerical implementation employing an explicit integration scheme. Of course, in a simple problem like the one at hand, this can be addressed by the adoption of implicit integrators.  Such a strategy, however, might be out of reach in more complex problems like studies of systems governed by beyond-GR theories.

\begin{figure}[t]
\centering
\includegraphics[width=9.2cm]{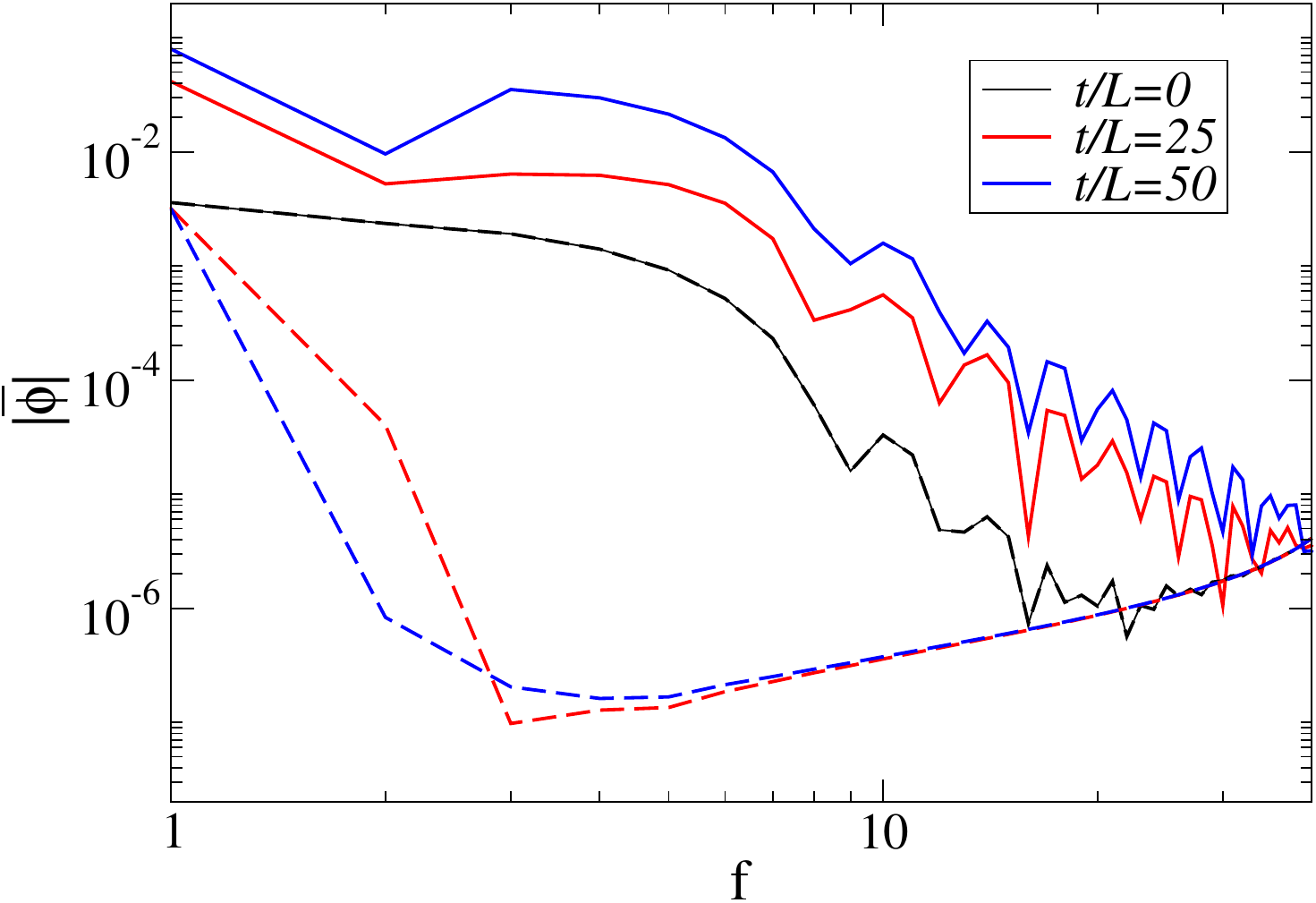}
\caption{Magnitude of the spatial Fourier transform of $\phi$ at fixed time, as a function of the mode number $f$ (wavenumber $k=2\pi f/L$), at times $t/L=0,25,50$. Solid lines correspond to the case
$\tau=0$ while dashed lines correspond to $\tau=1000$. As time progresses, a gradual increase in the energy content is observed for the case with no dissipation, while the opposite occurs when dissipation is turned on.}
\label{fourier_time}
\end{figure}

As a final illustration, Fig.~\ref{fourier_time} shows the spatial spectrum of $\phi$ at three times, for $\tau=0$ and $\tau=1000$, computed at the higher resolution $\Delta x= 10/4800$. Both problems are well posed, but, as time progresses, we see two different pictures. With dissipation, the energy in short-wavelength modes decreases and the solution stays within the EFT regime. Without dissipation, the opposite occurs, and the EFT regime is eventually abandoned, though with a ``rate of departure'' bounded by virtue of the problem being well-posed (see also~\cite{Figueras:2025gal}).
 
\section{Conclusions}
\label{sec:conclusion}
We have described a method that enhances the approach
presented in~\cite{Figueras:2024bba}. As shown, for sufficiently small $\{\gamma,\alpha\}$ couplings
the method~\cite{Figueras:2024bba,Figueras:2025gal} yields solutions that are consistent with the expected behavior. However, as their strengths are increased in tandem, the obtained solutions can be significantly different from the true one. This arguably undesirable behavior can be improved by choosing the coupling strength $\alpha$ of the added redundant operator several orders of magnitude smaller than $\gamma$, i.e.\ suppressing the redundant operator by a heavier scale. However, this carries the risk of inducing stiffness in the equation of motion and a consequent increase in the cost of numerical implementations.
Interestingly, the addition of dissipation helps to obtain a more consistent solution while reducing the risk of such an issue. Further, it provides another control knob to lessen the potential influence of short-wavelength modes that are aliased to long-wavelength ones at the numerical level in non-linear scenarios. 

The advantage of our approach is its  action-based formulation. Variational integrators discretize the action rather than the equations of motion, with the benefit of preserving the underlying symplectic structure over long evolutions~\cite{MarsdenWest2001,Tsang2015}. The boundary-plus-bulk action is a natural starting point for such schemes, where one evolves the bulk field directly, instead of imposing the outgoing solution analytically. Furthermore, whereas in this work the dissipation was introduced as a tool for fixing the classical conservative dynamics, one can envision using the bulk model for local, action-level implementations of real dissipation in settings where it is presently encoded differently -- for instance through the imaginary part of the in-out effective action in the world-line description of black-hole absorption~\cite{Goldberger:2005cd}.

In this work we focused on a simple scalar theory and briefly discussed the generalization to gravity, illustrating that the underlying formal approach can similarly be exploited to introduce a high-frequency damping mechanism. We leave a thorough study of the generalization to gravity for future work. A particular motivation for such a construction would be to explore the approach~\cite{Figueras:2024bba} in the context of a scenario that breaks the EFT regime and the consequences of the resulting UV control. As illustrated in~\cite{Franchini:2022ukz} for the formation of scalarized black holes in a specific beyond-GR theory (scalar Gauss--Bonnet), initial data defined within the EFT regime can dynamically lead to a solution that abandons it as collapse takes place. However, the stationary black hole with the same total mass is within the EFT regime.
It would be interesting to assess whether the approach
in~\cite{Figueras:2024bba} with and without the extension described here can also bridge these two regimes.

%################################################################

%########################################

%%%%%%%%%%%%%%%%%%
\section*{Acknowledgments}
We thank F. Abalos, C. Burgess, P. Figueras, A. Kovacs, I. Rothstein, S. Sibiryakov
and Thomas Sotiriou for comments.
The work of AB, LL and JR is supported by the 
Natural Sciences and Engineering Research Council (NSERC) of Canada. LL is also
supported in part by the Simons Foundation through Award SFI-MPS-BH-00012593-12.
 LL also thanks financial support via the Carlo Fidani Rainer Weiss Chair at 
 Perimeter Institute and CIFAR. 
 This research was supported
in part by Perimeter Institute for Theoretical Physics.
Research at Perimeter Institute is supported in part by
the Government of Canada through the Department of
Innovation, Science and Economic Development and by
the Province of Ontario through the Ministry of Colleges
and Universities.

\appendix

\section{In-In formulation}
\label{appendix}
%%%%%%%%%%%%%%%%%%%%%%%%%%%%%%%
In this section, following closely \cite{Besharat2024}, we illustrate how one can arrive at the dissipative dynamics \eqref{eq of mo2} by formally coupling the system \eqref{eq:action} to an unphysical thermal bath and then integrating it out in the Schwinger-Keldysh formalism. We take the bath to be made of fields $\xi$ at high temperature $T = \beta^{-1}$ and we assume a weak coupling, with coupling constant $g$, of the form
\begin{equation}
\label{eq:coupling}
    V[\phi,\xi] = \int dt \int d\mathbf{x} \, g \,  \phi \, \xi \, .
\end{equation}
In general, the non-equilibrium dynamics for $\phi$ can be obtained from the generating functional with doubled fields \cite{Kamenev2023},
\begin{equation}
    Z[J_+,J_-] = \int D \phi_{+} D\phi_{-} D\xi_{+} D\xi_{-} e^{i S[\phi_{+}] + i S_{\xi}[\xi_{+}] + i V[\phi_{+},\xi_{+}] + i\phi_+ J_+ - (+ \rightarrow -)} \, .
\end{equation}
Assuming that initially the $\phi$ and $\xi$ are not correlated, one can integrate out the fields $\xi$,
\begin{equation}
    Z[J_+, J_-] = \int D \phi_{+} D \phi_{-} e^{i S_{\text{eff}}[\phi_{+},\phi_{-}]} \, .
\end{equation}
At weak coupling \eqref{eq:coupling} the influence functional $\mathcal{I}[\phi_{+},\phi_{-}]$, defined by
\begin{equation}
\label{eq:Seff}
    S_{\text{eff}}[\phi_{+},\phi_{-}] \equiv S[\phi_+] - S[\phi_-] + \mathcal{I}[\phi_{+},\phi_{-}] \, ,
\end{equation}
is given by the \textit{linear response} approximation \cite{Besharat2023},
\begin{align}
\label{eq: SKI2}
{\cal I}=&\frac{ig^2}{2}\int dx_1 dx_2 \bigg\{
\phi_{+}(x_1) \,  \phi_{+}(x_2) \, G(x_1-x_2) + \phi_{-}(x_1) \, \phi_{-}(x_2) \, \tilde{G}(x_1-x_2)
 \nonumber \\
&~~~~~~~~~~~~~~~-\phi_{-}(x_1)\, \phi_{+}(x_2) \, K_{}(x_1-x_2) - \phi_{+}(x_1) \,  \phi_{-}(x_2) \, \tilde{K}(x_1-x_2) \bigg\}\;.
\end{align}
Here, $G$ and $\tilde{G}$ are the time-ordered and anti-time-ordered Green's functions,
\begin{align}
\label{TGereen}
G(x_1-x_2) &= \langle {\cal T} \big( \xi(x_1) \xi(x_2) \big) \rangle_{\beta} \, , \nonumber \\
\tilde{G}(x_1-x_2) &= \langle \tilde{\cal T} \big( \xi(x_1) \xi(x_2) \big) \rangle_{\beta}\;,
\end{align}
and $K$, $\tilde{K}$ are the unordered correlators,
\bseq
\label{GreenK}
\begin{align}
\label{GreenK1}
K(x_1-x_2) &= \langle \xi(x_1) \xi(x_2) \rangle_{\beta}\;, \\
\label{GreenK2}
\tilde{K}(x_1-x_2) &= \langle \xi(x_2) \xi(x_1) \rangle_{\beta}\; ,
\end{align}
\eseq
with expectation values taken at temperature $T = \beta^{-1}$.
It is convenient to perform a spatial Fourier transform. In particular,
\begin{equation}
     K(x)\equiv K(t,\mathbf{x}) = \int d \mathbf{k} \, e^{-i \mathbf{k} \mathbf{x}} \, K^{\mathbf{k}}(t) \, .
\end{equation}
The response of a \textit{thermal} bath is determined from the \textit{spectral density} defined via
\begin{align}
K^{\mathbf{k}}(t)=\int d\omega ~\e^{i\omega t }\rho^{\mathbf{k}}(\omega)\;.
\end{align}
More generally, for analytic $\rho$ we have \cite{Besharat2024}
\begin{equation}
\label{rho1HD}
\rho^{\mathbf{k}}(\omega)=\frac{1}{\beta}\rho^{\mathbf{k}}_{0}-\frac{\omega}{2}\rho^{\mathbf{k}}_{0}
+\frac{\omega^2}{\beta\Lambda^2}\bigg(\rho^{\mathbf{k}}_{2}+\frac{(\beta\Lambda)^2}{12}
\rho^{\mathbf{k}}_{0}\bigg)-\frac{\omega^3}{2\Lambda^2}\rho^{\mathbf{k}}_{2}+\ldots\; .
\end{equation}
with a scale $\Lambda$ that physically determines the retardation time of the bath. This expansion translates into the ultralocal expansions of all the necessary Green's functions,
\bseq
\label{highTGreens}
\begin{align}
\label{highTK}
&K^{\mathbf{k}}(t)=\frac{2\pi}{\beta}\rho^{\mathbf{k}}_{0}\,\delta(t)
+\pi i \rho^{\mathbf{k}}_{0}\,\delta'(t)-\frac{2\pi}{\beta \Lambda^2}
\check\rho^{\mathbf{k}}_{2}\,\delta''(t)
-\frac{i\pi}{\Lambda^2} \rho^{\mathbf{k}}_{2}\,\delta'''(t) + \dots\;,\\
\label{highTtK}
&\tilde{K}^{\mathbf{k}}(t)=\frac{2\pi}{\beta}\rho^{\mathbf{k}}_{0}\,\delta(t)
-\pi i \rho^{\mathbf{k}}_{0}\,\delta'(t)-\frac{2\pi}{\beta \Lambda^2}
\check\rho^{\mathbf{k}}_{2}\,\delta''(t)
+\frac{i\pi}{\Lambda^2} \rho^{\mathbf{k}}_{2}\,\delta'''(t) + \dots\;,\\
\label{highTG}
&G^{\mathbf{k}}(t)=\frac{2\pi}{\beta}\rho^{\mathbf{k}}_{0}\,\delta(t)
+\pi i \rho^{\mathbf{k}}_{0}\, \text{sign}(t) \delta'(t) - \frac{2\pi}{\beta \Lambda^2}
\check\rho^{\mathbf{k}}_{2}\,\delta''(t)
-\frac{i\pi}{\Lambda^2} \rho^{\mathbf{k}}_{2} \, \text{sign}(t)
\delta'''(t) + \dots\;,\\
\label{highTtG}
&\tilde {G}^{\mathbf{k}}(t)=\frac{2\pi}{\beta}\rho^{\mathbf{k}}_{0}\,\delta(t)
-\pi i \rho^{\mathbf{k}}_{0}\,\text{sign}(t)\delta'(t) -\frac{2\pi}{\beta \Lambda^2}
\check\rho^{\mathbf{k}}_{2}\,\delta''(t)
+\frac{i\pi}{\Lambda^2} \rho^{\mathbf{k}}_{2} \, \text{sign}(t)
\delta'''(t) + \dots \;,
\end{align}
\eseq
where we have introduced
\begin{equation}
\label{rhocheck}
\check \rho^{\mathbf{k}}_{2}=\rho^{\mathbf{k}}_{2}+\frac{(\beta \Lambda)^2}{12}
\rho^{\mathbf{k}}_{0} \, .
\end{equation}
Straightforward calculation gives \footnote{Here we have repeatedly used $\text{sign}(t) = \theta(t) - \theta(-t)$, and the identity
\begin{align}
\int dt~ \theta(t) f(t) \delta'(t)=-\frac{1}{2}f'(0) - \delta(0) f(0)\;.
\end{align}
The terms with $\delta(0)$ simply renormalize the (doubled) action of the system and so we drop them.}
\begin{equation}
    \mathcal{I} = \pi i g^2 \int dt \int_{\mathbf{k}} \frac{4}{\beta} \, \rho_{0}^{\mathbf{k}} \,  \hat{\phi}_{-\mathbf{k}} \hat{\phi}_{\mathbf{k}} + 2i\rho_{0}^{\mathbf{k}} \dot{\bar{\phi}}_{- \mathbf{k}} \hat{\phi}_{\mathbf{k}} - \frac{4}{\beta} \frac{\rho_{2}^{\mathbf{k}}}{\Lambda^2} \partial_t^2 \big(\hat{\phi}_{-\mathbf{k}}\big) \hat{\phi}_{\mathbf{k}}  -2i\frac{\rho_{2}^{\mathbf{k}}}{\Lambda^2} \big(\partial_t^2\dot{\bar{\phi}}_{- \mathbf{k}} \big) \hat{\phi}_{\mathbf{k}} + \dots \, ,
\end{equation}
where we have introduced the classical and quantum fields,
\begin{equation}
\label{eq:phicq}
    \phi_{\pm} \equiv \bar{\phi} \pm \hat{\phi} \, .
\end{equation}
Now, to get the desired response, we fix the spectral density to be of the form
\begin{equation}
    \rho^{\mathbf{k}}_0 = \rho_0 \frac{\mathbf{k}^2}{\Lambda^2} \, ,  \quad \rho_2^{\mathbf{k}} = - \rho_0 \, ,
\end{equation}
with $\rho_0 > 0$, and all higher expansion coefficients set to zero. Note that this implies that the spectral density is \textit{not} positive definite for $\omega^2 > \mathbf{k}^2$, which precludes a physical interpretation of this bath. Such a choice of the spectral density is physically equivalent to the choice of the bulk action with higher-derivative terms \eqref{action+}. Neglecting the terms suppressed in the high-temperature limit, $\beta \Lambda \ll 1$, we have
\begin{equation}
    \check \rho_2^{\mathbf{k}} \simeq - \rho_0 \, .
\end{equation}
Going back to the $\mathbf{x}$-space we finally arrive at
\begin{equation}
     \mathcal{I} = 2 i \tau \int dt \int d \mathbf{x} \Bigg[ -\frac{2}{\beta} \hat{\phi}\Box\hat{\phi} - i \hat{\phi} \partial_t \Box \bar{\phi} \Bigg]  \, ,
\end{equation}
with $\tau \equiv \frac{\pi g^2 \rho_0}{\Lambda^2}$. Due to the sign-indefiniteness of the spectral density, not all the configurations of $\hat{\phi}$ are suppressed in the high-temperature limit. \footnote{In the path integral we get a weight
\begin{equation}
     e^{i \mathcal{I}} =  e^{ - 4 \tau/\beta \left[\int (\partial_{\mathbf{x}} \hat{\phi})^2 - \int (\partial_{t} \hat{\phi})^2 \right]} \, .
\end{equation}
Note that the second term displays the growing directions explicitly} The particular structure of this influence functional is fixed completely by the fluctuation-dissipation relations, or KMS symmetry, and so if we want to have a dissipative term of the form $\hat{\phi} \partial_t \Box \bar{\phi}$, it must be accompanied by the fluctuation term $\frac{2}{\beta} \hat{\phi}\Box\hat{\phi}$. This means that the absence of the uniform suppression of $\hat{\phi}$'s is inevitable.

We then expand
\begin{equation}
\label{eq:actiondiff}
    S[\phi_+] - S[\phi_-]  = 2\int dt \int d \mathbf{x} \, \frac{\delta S}{ \delta \bar{\phi}} \hat{\phi} + \dots \, ,
\end{equation}
The $\hat\phi$ integral is understood as a formal Gaussian integral, evaluated at its stationary point. After using the \textit{Hubbard--Stratonovich} transformation,
\begin{equation}
    e^{\frac{4\tau}{\beta} \hat{\phi}\, \Box\, \hat{\phi}} = \int D \eta \,  e^{\frac{\beta}{4\tau} \eta \,  \Box^{-1} \eta + 2i \eta \hat{\phi}} \, ,
\end{equation}
the full partition function becomes
\begin{equation}
\label{eq: partition function}
    Z = \int D \eta \, e^{\frac{\beta}{4\tau} \eta \,  \Box^{-1} \eta}\int D \hat{\phi} D \bar{\phi} \, e^{i S_{\text{eff}}} \, ,
\end{equation}
with
\begin{equation}
    S_{\text{eff}} = 2\int dt \int d \mathbf{x} \Bigg(\frac{\delta S}{ \delta \bar{\phi}} + \tau \partial_t \Box \bar{\phi} + \eta\Bigg) \hat{\phi} \, .
\end{equation}
Stationarity of $S_{\text{eff}}$ in $\hat\phi$ imposes Eq.~\eqref{eq of mo2} sourced by $\eta$. Setting the noise to zero yields precisely the sought-after dissipative equation of motion.
%%%%%%%%%%%%%%%%%%%%%%%%%%%%%%%

\printbibliography[heading=bibintoc]
\end{document}